\documentclass{ifacconf}

\usepackage{graphicx}      
\usepackage{natbib}        

\usepackage{microtype} 
\usepackage{amsmath,amssymb} 
\usepackage{supertabular,booktabs} 
\usepackage{url}
\usepackage{tikz} 
\usetikzlibrary{shapes} 

\begin{document}
\begin{frontmatter}

\title{A Vehicle Routing Problem for Human-Centered Electric Mobility}

\thanks[footnoteinfo]{This work is part of [project MORE -- Munich Mobility Research Campus] and funded by dtec.bw -- Digitalization and Technology Research Center of the Bundeswehr. dtec.bw is funded by the European Union -- NextGenerationEU.}

\author[First]{Mostafa Emam,} 
\author[First]{Bj\"{o}rn Martens,} 
\author[First]{Thomas Rottmann,}
\author[First]{Matthias Gerdts}

\address[First]{Institute of Applied Mathematics and Scientific Computing, University of the Bundeswehr Munich
	(e-mail: mostafa.emam@unibw.de,
	bjoern.martens@unibw.de,
	thomas.rottmann@unibw.de,
	matthias.gerdts@unibw.de)}

\begin{abstract}
	In this paper, we present the Electric Mobility Dial-a-Ride Problem (EM-DARP), which extends the Electric Vehicle Dial-a-Ride Problem (EV-DARP) to better accommodate human-focused mobility services. The problem involves utilizing a fleet of heterogeneous Electric Vehicles (EVs) to fulfill a set of customer requests with DARP and mobility-related specifications, while incorporating visits to charging stations amid requests. The problem is formulated as a Mixed-Integer Linear Program (MILP) and subsequently solved for a number of curated evaluation scenarios to demonstrate its practical applicability.
\end{abstract}


\end{frontmatter}

%
%
%

\section{Introduction}
Environmental aspects, government regulations, and cost reduction have been the driving factors in pursuing \textit{Green Transport} with low-to-zero carbon emissions in urban and intercity transportation. In this context, the EV-DARP excels for on-demand services, combining typical routing elements, such as time windows and pickup and delivery constraints, with the battery limitations of EVs. However, the EV-DARP does not fully represent modern shuttles, which usually have adjustable interiors to accommodate mobility equipment of the elderly and passengers with disabilities. To the best of our knowledge, integrating the EV-DARP and vehicles with configurable capacities has not yet been established, which is what we pursue in this work. Given its role in on-demand mobility, we denote the problem \textit{Electric Mobility DARP} (EM-DARP) and explain its attributes and MILP formulation in the sequel. We conclude with a small-scale simulation study, and plan to explore more efficient solution methods in future work.

%
%
%

\section{Literature Review}
Ever since its inception by \cite{c0108}, the Vehicle Routing Problem (VRP) and its variants have been extensively investigated in combinatorial optimization and operations research. For instance, \cite{c0109} studied the Pickup and Delivery Problem (PDP), which is a VRP variant with precedence constraints. In the VRPTW, \cite{c0093} suggested incorporating Time Window (TW) restrictions on the customer service times, and later collaborated with \cite{c0085} to examine additional solution methods and acceleration strategies. \cite{c0087} conducted a survey on DARP methods, which generalize the PDP and VRPTW while taking the human element into consideration. With the rising popularity of EVs, \cite{c0086} introduced the E-VRPTW, which employs EVs to service customers and accordingly integrates visits to charging stations within the travel routes. More recently, \cite{c0088} studied handling requests with soft time windows and probabilistic waiting times at recharging stations, while \cite{c0082} investigated utilizing Battery Swapping Stations (BSS) with constant service times in lieu of charging stations. \cite{c0089} analyzed the impact of partial recharging and the application of linear and nonlinear charging models on travel routes and battery life. Similarly, \cite{c0099} explored nonlinear charging and load-dependent discharging. In terms of mobility, most studies handle different kinds of passengers, e.g., with or without mobility equipment, by either associating them with specific agents or by utilizing disjunct decision variables. \cite{c0104} adopted the latter method, in which vehicles have two separate variables for tracking the number of impaired and unimpaired patients. Similar examples can be found in the survey by \cite{c0087}. Conversely, \cite{c0106} introduced the concept of configurable capacity, which enables a vehicle to transport a varying number of passengers and support equipment based on its initial configuration. \cite{c0107} later extended this approach by allowing vehicles to modify their configuration mid-route, thereby generating more realistic and optimized routes.

\begin{figure}[htp]
	\centering
	\includegraphics[width=0.135\textwidth]{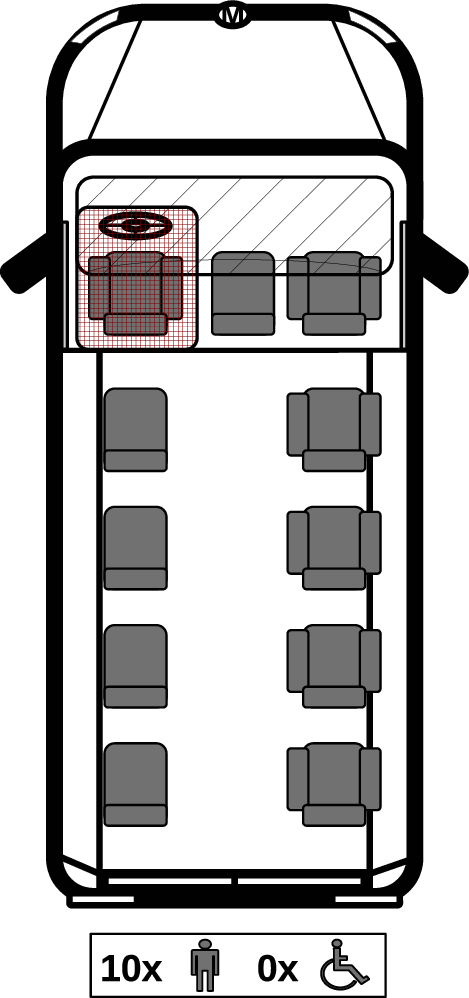}\hspace{10pt}
	\includegraphics[width=0.135\textwidth]{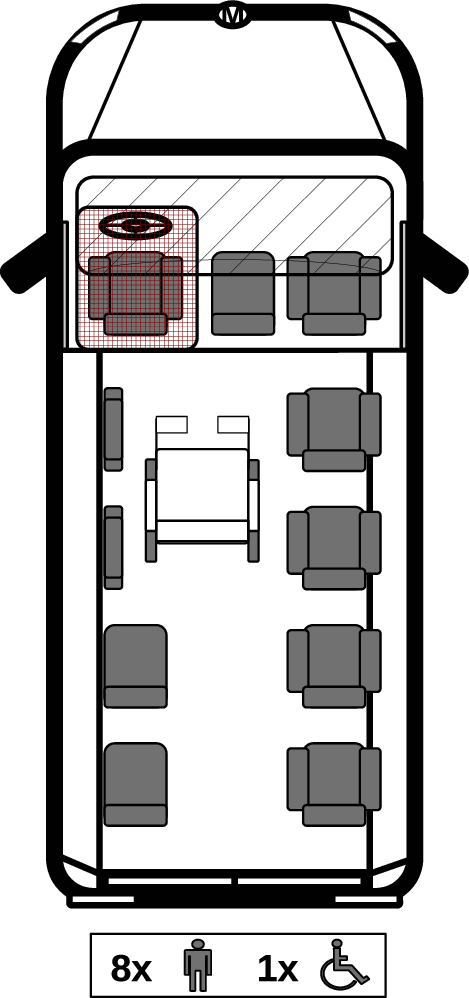}\hspace{10pt}
	\includegraphics[width=0.135\textwidth]{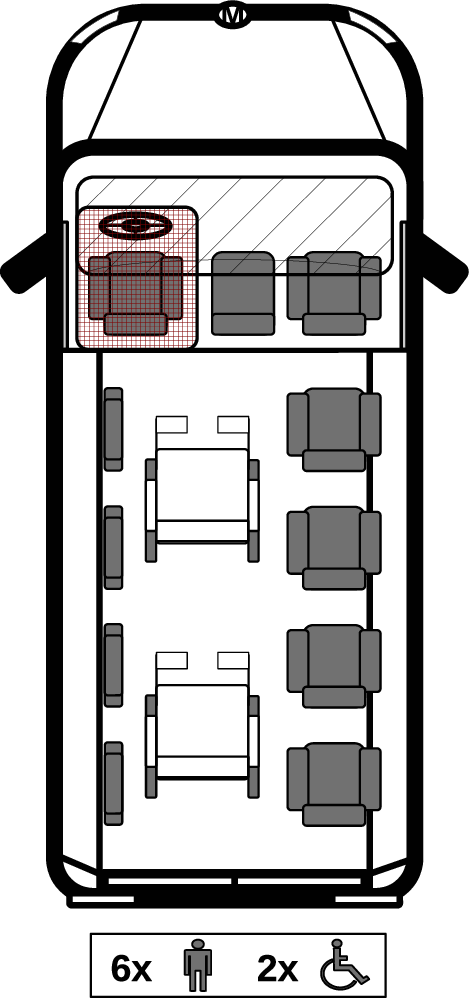}
	\caption{EM-DARP shuttle with three possible configurations}
	\label{fig:shuttle_configs}
\end{figure}

We combine these concepts to design the EM-DARP, which includes five core elements: (i) PDP and (ii) TW constraints, as well as a (iii) battery model outlining the discharging and recharging behaviors, whether through simple or complex relationships. Here, we extend the notion of rechargeability to denote the agent's capability to increase its energy or State-of-Charge (SoC) level and, subsequently, its driving time. For instance, energy can be replenished at a charging station, a BSS, or using on-road wireless charging, cf. \cite{c0105}. Furthermore, we incorporate (iv) modular or configurable capacity, in which an agent can modify its configuration to accommodate different passenger types, as depicted in Fig.~\ref{fig:shuttle_configs}. Finally, there exists a parameter or method to differentiate requests based on their (v) level of importance or criticality; for instance, to prioritize requests involving mobility equipment over others.

%
%
%

\section{Mathematical Formulation}
In this section, we mathematically define the EM-DARP and include additional elements to address more realistic routing problems. Here, we present a multi-agent, multi-depot model that incorporates heterogeneous agents and requests. It employs soft TW restrictions and a selective variant that permits the rejection of requests. Furthermore, it supports Open VRPs (OVRP), where agents are not required to return to a depot after route completion. For the battery model, we propose load-dependent discharging and nonlinear charging with partial recharge capabilities. Moreover, we use the term \textit{charging station} to denote a charging point that can service one agent at most, and thereby introduce precedence constraints to ensure that vehicles can visit them multiple times, but only sequentially and not concurrently. We highlight that this approach can be readily extended to support multiple charging points at the same node by adding a number of virtual \textit{stations} at the same position, equal to the number of charging points.

%
%
\subsection{Model Inputs}

Our problem has three types of inputs: the road network, the customer requests, and the fleet agents. First, we model the network as a directed graph $\mathcal{G} = (\mathcal{V}, \mathcal{E})$, where the nodes $\mathcal{V}$ represent physical locations, and the edges $\mathcal{E}$ are the travel arcs connecting them. The nodes include the agents' starting positions $\mathcal{H}_0$, and the pickup $\mathcal{L}^p$ and delivery $\mathcal{L}^d$ locations, with the set $\mathcal{L} = \mathcal{L}^p \cup \mathcal{L}^d$. In addition, we have the set of charging stations $\mathcal{F}$, which contains duplicates to enable multiple visits to the same station. We model this as $\mathcal{F} = \bigcup_{j = 0}^n \{\widetilde{\mathcal{F}}^j\}$, with $\widetilde{\mathcal{F}}^j = \bigcup_{i = 0}^{m-1} \{f_i\}$, where $m$ is the number of stations, $n$ is the number of additional visits per station, and $n=0$ signifies a single visit per station. For example, $f_0^0$ represents the first visit to the first station and $f_0^1$ is the second visit to the first station. Since the number of constraints increases with $n$, its value is a compromise between flexibility and problem complexity. Lastly, we have the final depots $\mathcal{H}_f$, which yields the full set $\mathcal{V} = \mathcal{H}_0 \cup \mathcal{L} \cup \mathcal{F} \cup \mathcal{H}_f$. Each arc in $\mathcal{E}$ is usually associated with a travel distance, time, and cost. Here, we adopt the simplification proposed by \cite{c0086}, i.e., assume an average travel speed of $1~[ms^{-1}]$ and compute the arc cost in terms of travel time. This yields the cost $c_{ij} > 0, \forall i,j \in \mathcal{V}, i \neq j$, with $c_{ii} = 0$, which interchangeably represents cost, time, and distance. Note that this approach also supports asymmetric weights with $c_{ij} \neq c_{ji}, \forall i,j \in \mathcal{V}$.

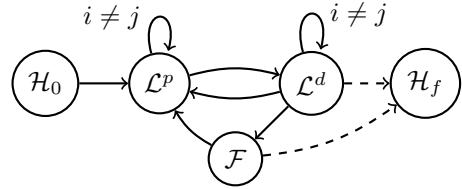
\begin{figure}[htb]
	\centering
	\begin{tikzpicture}[->]
	\begin{scope}[every node/.style={draw=black,thick}]]
	\node[circle,draw] (H0) at (0.5,0) {$\mathcal{H}_0$};
	\node[circle,draw] (Lp) at (2,0) {$\mathcal{L}^p$};
	\node[circle,draw] (Ld) at (4,0) {$\mathcal{L}^d$};
	\node[circle,draw] (Hf) at (5.5,0) {$\mathcal{H}_f$};
	\node[circle,draw] (S) at (3,-1) {$\mathcal{F}$};
	\end{scope}
	\begin{scope}[every edge/.style={draw=black,thick}]]
	\path (H0) edge node{}(Lp);
	\path (Lp) edge [loop above] node[left]{$i \neq j~$}(Lp);
	\path (Lp) edge [bend left=15] node{}(Ld);
	\path (Ld) edge [loop above] node[right]{$~i \neq j$}(Ld);
	\path (Ld) edge [bend left=15] node{}(Lp);
	\path (Ld) edge node{}(S);
	\path (Ld) edge [dashed]node{}(Hf);
	\path (S) edge [bend left=15] node{}(Lp);
	\path (S) edge [dashed,bend right=15] node{}(Hf);
	\end{scope}
	\end{tikzpicture}
	\caption{The possible travel routes between all nodes in $\mathcal{V}$}
	\label{fig:network_routes}
\end{figure}

Agents traverse the road network as shown in Fig. \ref{fig:network_routes}. For instance, an agent may only leave its initial node in $\mathcal{H}_0$ to a pickup node in $\mathcal{L}^p$. Afterwards, it either visits a different pickup, denoted by the loop with the condition $i \neq j$, or travels to a delivery node in $\mathcal{L}^d$. This strategy significantly reduces the number of binary variables, thereby elevating the model efficiency. The dashed lines from $\mathcal{L}^d$ and $\mathcal{F}$ to $\mathcal{H}_f$ signify the difference between closed and open VRPs. For the former, agents must terminate their route at a depot in $\mathcal{H}_f$; for the latter, an agent will end its route either after completing its last delivery in $\mathcal{L}^d$ or at a station in $\mathcal{F}$.

\begin{table}[htb]
	\caption{Request parameters}
	\label{tbl:request_params}
	\begin{supertabular*}{0.48\textwidth}{p{0.08\textwidth} p{0.35\textwidth}}
		Symbol & Description\\
		\midrule
		$p^r \in \mathcal{L}^p$ & pickup location\vspace*{2pt}\\
		$d^r \in \mathcal{L}^d$ & delivery (drop-off) location\\
		$q_1^r \in \mathbb{N}$ & number of passengers\\
		$q_2^r \in \mathbb{N}_0$ & number of mobility equipment (e.g., wheelchairs)\\
		$s^r \ge 0$ & total request service time\\
		$\varrho^r \in \{p, d\}$ & TW restriction type: $p$ for pickup, $d$ for delivery\\
		$\underline{t}^r \ge 0$ & lower bound on the request time window (TW)\\
		$\overline{t}^r > \underline{t}^r$ & upper bound on the request time window (TW)\\
		$\lambda^r \ge 1$ & request priority\\
		\bottomrule
	\end{supertabular*}
\end{table}

Second, each request $r \in \mathcal{R}$ has the parameters in table \ref{tbl:request_params}. Since $\mathcal{L}$ comprises request pickup and delivery locations, we introduce the mapping function $\Gamma: \mathcal{L} \rightarrow \mathcal{R}$, which associates locations with their corresponding requests. For simplicity, we assume that all equipment types occupy the same space and are handled uniformly. $s^r$ includes the time required to modify the agent's configuration, which takes a few seconds in the case of foldable seats, as mentioned by \cite{c0107}. Finally, larger values of $\lambda^r \gg 1$ signify a high request priority, thereby heavily penalizing TW violations and reducing the likelihood of request rejection.

\begin{table}[htb]
	\caption{Agent parameters}
	\label{tbl:agent_params}
	\begin{supertabular*}{0.48\textwidth}{p{0.09\textwidth} p{0.35\textwidth}}
		Symbol & Description\\
		\midrule
		$v^k \in \mathcal{H}_0$ & initial node (hub, location, or charging station)\\
		$\delta^k \ge 0$ & initial delay, time of earliest availability at $v^k$\vspace*{4pt}\\
		$Q_1^k \in \mathbb{N}$ & maximum passenger load (capacity)\vspace*{2pt}\\
		$Q_2^k \in \mathbb{N}$ & maximum equipment load (capacity)\vspace*{2pt}\\
		$\gamma^k \ge 1$ & equipment-to-passenger conversion factor\vspace*{2pt}\\
		$D^k \ge 0$ & remaining trip duration, e.g., workday\\
		$S^k \ge 0$ & required service time at charging stations\\
		$\underline{\sigma}^k \in (0,1]$ & minimum operational SoC\\
		$\sigma^k \in [\underline{\sigma}^k, 1]$ & initial SoC at $v^k$\\
		$\widetilde{\sigma}^k \in [\underline{\sigma}^k, 1]$ & minimum desired SoC upon leaving a station $\in \mathcal{F}$\\
		\bottomrule
	\end{supertabular*}
\end{table}

Third, each agent $k \in \mathcal{K}$ has the parameters in table \ref{tbl:agent_params}. $\delta^k > 0$ accommodates agents with delayed availability, e.g., due to charging or completing a previous route. In Section \ref{ssb:cstr_capacity}, we explain how $\gamma^k$ is used to model the configurable capacity. To increase autonomy while preventing battery degradation, \cite{c0089} recommend using the SoC limits $\underline{\sigma}^k = 0.25$ and $\widetilde{\sigma}^k = 0.85$. We also highlight that higher values of $\widetilde{\sigma}^k$ discourage redundant visits to $\mathcal{F}$.

\begin{table}[htb]
	\caption{Battery model parameters}
	\label{tbl:battery_params}
	\begin{supertabular*}{0.48\textwidth}{p{0.07\textwidth} p{0.35\textwidth}}
		Symbol & Description\\
		\midrule
		$\alpha_0 > 0$ & default discharge rate (no passengers on board)\\
		$\alpha_1 > 0$ & load-dependent (passengers) discharge rate\\
		$\alpha_2 > 0$ & load-dependent (equipment) discharge rate\\
		$\beta_1 > 0$ & charging rate within the segment $\sigma^k \in [0.00,\:0.85]$\\
		$\beta_2 > 0$ & charging rate within the segment $\sigma^k \in (0.85,\:0.95]$\\
		$\beta_3 > 0$ & charging rate within the segment $\sigma^k \in (0.95,\:1.00]$\\
		\bottomrule
	\end{supertabular*}
\end{table}

\begin{figure}[htb]
\centering
\includegraphics[width=0.42\textwidth]{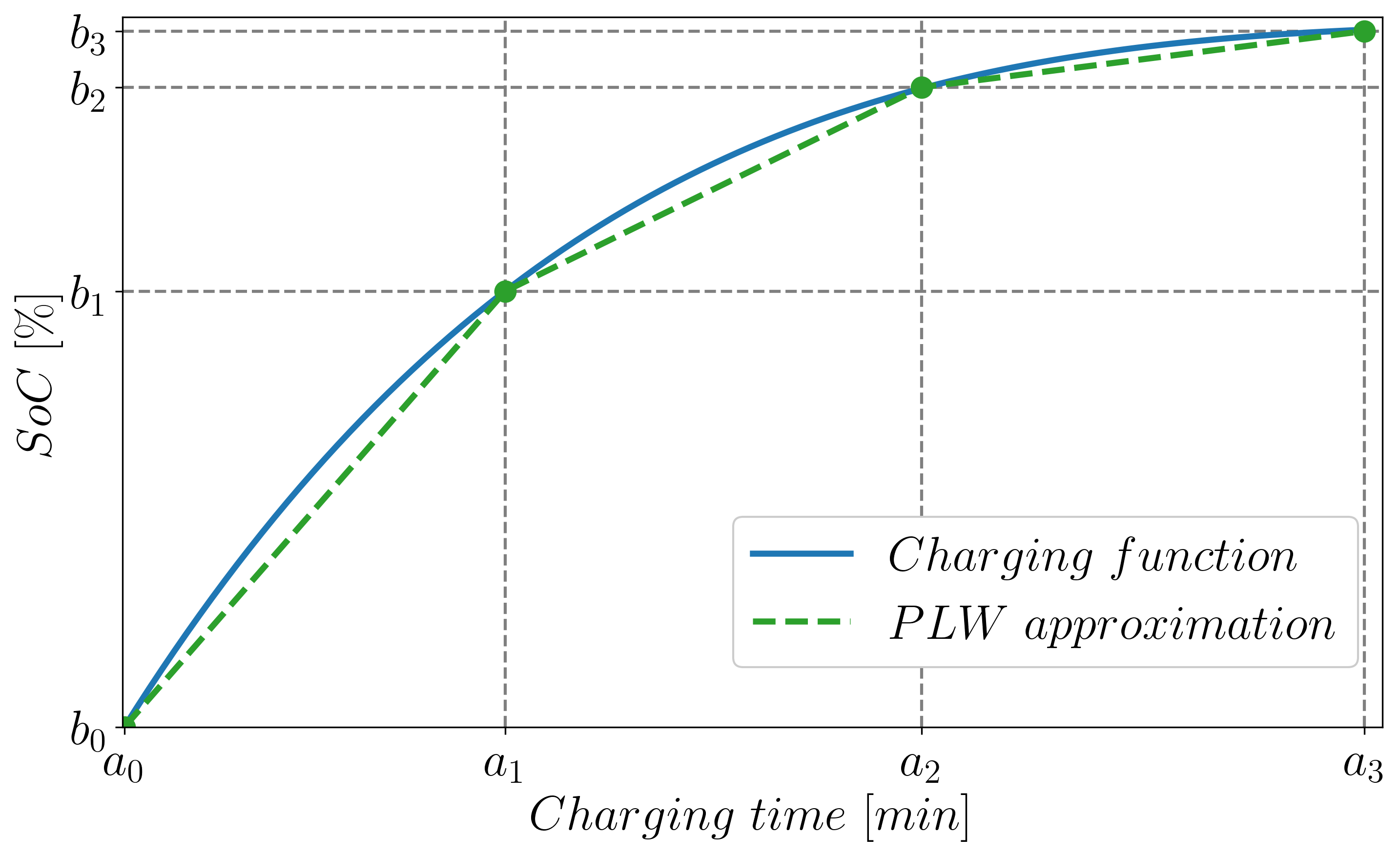}
\caption{Nonlinear charging function and its approximation}
\label{fig:battery_charge_fcn}
\end{figure}

For simplicity, we assume that all agents have the same battery, charger type, and charging protocol, resulting in a unified discharging and charging behavior. We propose a battery model with linear, load-dependent discharging and a piecewise linear (PLW) approximation of the nonlinear charging function shown in Fig. \ref{fig:battery_charge_fcn}. We highlight that the PLW approximation forms a connected region with strictly decreasing gradients $\beta_3 < \beta_2 < \beta_1$, such that by modeling each segment with a linear function $g_i(x) = a_i + b_i x$, the desired value becomes the lower envelope of the set of functions $\{g_1, g_2, g_3\}$. In this case, \cite{c0103} proposes an efficient modeling using only continuous variables in a type 2 Specially Ordered Set (SOS2). However, we opt to use binary variables to extend the applicability of our approach to any MILP solving tool. Finally, we present the model sets, parameters, and decision variables in table \ref{tbl:milp_model_sets_params_vars}.

\begin{table}[htb]
\caption{Model sets, parameters, and variables}
\label{tbl:milp_model_sets_params_vars}
\begin{supertabular*}{0.48\textwidth}{p{0.086\textwidth} p{0.36\textwidth}}
	Symbol & Description\\
	\midrule
	$\mathcal{R}$ & set of passenger requests\\
	$\mathcal{K}$ & set of heterogeneous electric vehicles (agents)\\
	$\mathcal{H}_0$ & set of initial agent nodes $\mathcal{H}_0 = \bigcup_{k \in \mathcal{K}} \{v^k\}$\vspace*{4pt}\\
	$\mathcal{L}^p$ & set of pickup locations $\mathcal{L}^p=\bigcup_{r\in \mathcal{R}} \{p^r\}$\vspace*{2pt}\\
	$\mathcal{L}^d$ & set of delivery locations $\mathcal{L}^d=\bigcup_{r\in \mathcal{R}} \{d^r\}$\vspace*{2pt}\\	
	$\mathcal{L}$ & set of pickup and delivery locations $\mathcal{L} =\mathcal{L}^p \cup \mathcal{L}^d$\\
	$\mathcal{F}$ & set of charging stations with duplicates\\
	& $\mathcal{F} =\bigcup_{i\in \{0,1,\ldots,m-1\},\:j \in \{0,1,\ldots,n\}} \{f_i^j\}$\vspace*{2pt}\\
	$\mathcal{H}_f$ & set of final depot (hub) nodes\\
	$\mathcal{V}$ & complete set of all nodes $\mathcal{V}= \mathcal{H}_0 \cup \mathcal{L} \cup \mathcal{F} \cup \mathcal{H}_f$\\
	\midrule
	$c_{ij} \ge 0$ & travel time between nodes $i, j \in \mathcal{V}$, with $c_{ii} = 0$\\
	$\mu_i \in \{$-$1,1\}$ & load rectifying factor, $1$ if $i \in \mathcal{L}^p$; $-1$ if $i \in \mathcal{L}^d$ \\
	$m \in \mathbb{N}$ & number of charging stations\\
	$n \in \mathbb{N}_0$ & number of charging stations duplicates, which is also the number of additional visits per station\\
	$\epsilon > 0$ & small objective weight\\
	$\zeta > 0$ & medium objective weight, with $\zeta > \epsilon$\\
	$\eta \gg 0$ & large objective weight, with $\eta > \zeta$\\
	$M \gg 1$ & sufficiently large value for big-M constraints\\
	\midrule
	$x_{ij}^k \in \{0, 1\}$ & 1 if agent $k \in \mathcal{K}$ travels from node $i \in \mathcal{V}$ to node $j \in \mathcal{V}$; 0 otherwise\\
	$y_r \in \{0, 1\}$ & 1 if request $r \in \mathcal{R}$ is accepted (assigned); 0 otherwise\\
	$T^k \ge 0$ & agent trip duration\\
	$T \ge 0$ & mission duration (maximum trip time of all agents)\\
	$t_i \ge 0$ & arrival time (begin of service) at node $i \in \mathcal{L} \cup \mathcal{F}$\\
	$\tau_i \ge 0$ & slack for TW violation at node $i \in \mathcal{L}$\\
	$T_r \ge 0$ & auxiliary variable for arrival times of request $r \in \mathcal{R}$\\
	$\varDelta_r \ge 0$ & auxiliary variable for TW violation of request $r \in \mathcal{R}$\\
	$u_{i,1}^k \in \mathbb{N}_0$ & passenger load of agent $k \in \mathcal{K}$ on leaving $i \in \mathcal{L}$\\
	$u_{i,2}^k \in \mathbb{N}_0$ & equipment load of agent $k \in \mathcal{K}$ on leaving $i \in \mathcal{L}$\\
	$\phi_i^k \in [0,1]$ & SoC of agent $k \in \mathcal{K}$ on arriving at node $i \in \mathcal{V} \setminus \mathcal{H}_0$\\
	$\xi_{i,1} \ge 0$ & charging time at station $i \in \mathcal{F}$ with $\beta_1$\\
	$\xi_{i,2} \ge 0$ & charging time at station $i \in \mathcal{F}$ with $\beta_2$\\
	$\xi_{i,3} \ge 0$ & charging time at station $i \in \mathcal{F}$ with $\beta_3$\\
	$z_{i,1} \in \{0,1\}$ & 1 if charging at station $i \in \mathcal{F}$ fully utilizes the first segment, i.e., if the final SoC $\ge 0.85$; 0 otherwise\\
	$z_{i,2} \in \{0,1\}$ & 1 if charging at station $i \in \mathcal{F}$ fully utilizes the second segment, i.e., if the final SoC $\ge 0.95$; 0 otherwise\\
	\bottomrule
\end{supertabular*}
\end{table}

%
%
\subsection{Objective Function}
The objective function \eqref{eqn:obj_fcn} aims to minimize the mission duration, arrival times and TW violations of accepted requests, and the number of rejected requests. It also manipulates the model to prioritize fulfilling certain requests over others based on their comparative values of $\lambda^r, \forall r \in \mathcal{R}$.
\begin{equation}
\min\quad T + \sum_{r \in \mathcal{R}} \lambda^r \left(\epsilon~T_r + \zeta~\varDelta_r + \eta~(1 - y_r)\right) \label{eqn:obj_fcn}
\end{equation}
When a request is accepted, it satisfies the condition $y_r = 1$ and has the arrival times $t_{p^r}$ and $t_{d^r}$; otherwise, all values shall be equal to zero. We model this dependency as a product of logical and continuous variables of the form $g \triangleq \delta f(x)$, with $g\ge 0, \:\delta \in \{0, 1\}, \:f(x) \in \mathbb{R}^+$. As stated by \cite{c0081}, we can linearize this product and substitute it with an auxiliary variable $T_r$, subject to:
\begin{align}
t_{p^r} + t_{d^r} - M (1 - y_r) \le &~T_r \le t_{p^r} + t_{d^r}, &&~\forall r \in \mathcal{R} \label{eqn:obj_time_lin_w1}\\
0 \le &~T_r \le M y_r, &&~\forall r \in \mathcal{R} \label{eqn:obj_time_lin_w2}
\end{align}

Similarly, we introduce the auxiliary slack variable $\varDelta_r$ for penalizing the TW violation of request $r$, subject to:
\begin{align}
\tau_{p^r} + \tau_{d^r} - M (1 - y_r) \le &~\varDelta_r \le \tau_{p^r} + \tau_{d^r}, &&~\forall r \in \mathcal{R} \label{eqn:obj_slack_lin_w1}\\
0 \le &~\varDelta_r \le M y_r, &&~\forall r \in \mathcal{R} \label{eqn:obj_slack_lin_w2}
\end{align}

We now delineate the rest of the model constraints.

%
%
\subsection{Request Assignment and Flow Conservation}
Equation \eqref{eqn:cstr_req_accept} permits acceptance and rejection of requests, in which adjusting it to $y_r = 1$ forces the acceptance of all requests, i.e., the non-selective problem variant. When a request is accepted, its pickup and delivery locations are visited in accordance with \eqref{eqn:cstr_req_enter_p} and \eqref{eqn:cstr_req_enter_d}, respectively. We guarantee that the same agent visits both locations with \eqref{eqn:cstr_same_agent_p_d}. Equation \eqref{eqn:cstr_chg_station_precedence} ensures that each charging station is visited by one agent at most and upholds the sequence of visits, such that a station is re-entered only if its predecessor has been visited. In other words, charging station duplicates $\widetilde{\mathcal{F}}^{j>0}$ are visited only after the original stations $\widetilde{\mathcal{F}}^0$ and in an increasing order of $j = \{1, \ldots, n\}$. For each agent, we limit the number of outgoing arcs from $\mathcal{H}_0$ to a single pickup with \eqref{eqn:cstr_agent_leave_hub}, while permitting idle agents to remain at their initial positions. Lastly, equations \eqref{eqn:cstr_flow_continuity_lp}, \eqref{eqn:cstr_flow_continuity_ld}, and \eqref{eqn:cstr_flow_continuity_f} guarantee flow continuity at $\mathcal{L}^p$, $\mathcal{L}^d$, and $\mathcal{F}$, respectively.
\begingroup
\allowdisplaybreaks
\begin{align}
&y_r \le 1, \hspace{160pt}\forall r \in \mathcal{R} \label{eqn:cstr_req_accept}\\
&y_r = \sum_{k \in \mathcal{K}} x_{v^k p^r}^k + \sum_{k \in \mathcal{K}}\sum_{i \in \mathcal{L} \cup \mathcal{F} \setminus\{p^r\}} x_{i p^r}^k, \hspace{29pt}\forall r \in \mathcal{R} \label{eqn:cstr_req_enter_p}\\
&y_r = \sum_{k \in \mathcal{K}} \sum_{i \in \mathcal{L} \setminus\{d^r\}} x_{i d^r}^k, \hspace{94pt}\forall r \in \mathcal{R} \label{eqn:cstr_req_enter_d}\\
\begin{split}
&x_{v^k p^r}^k + \sum_{i \in \mathcal{L} \cup \mathcal{F} \setminus\{p^r\}} x_{i p^r}^k = \sum_{i \in \mathcal{L} \setminus\{d^r\}} x_{i d^r}^k,\\
&\hspace{163pt}\forall r \in \mathcal{R}, k \in \mathcal{K}
\end{split}\label{eqn:cstr_same_agent_p_d}\\
\begin{split}
&\sum_{k \in \mathcal{K}} \sum_{h \in \mathcal{L}^d} x_{h f_i^j}^k \le \sum_{k \in \mathcal{K}} \sum_{h \in \mathcal{L}^d} x_{h f_i^{j-1}}^k \le 1, \\
&\hspace{52pt}\forall i = \{0, 1, \ldots, m-1\},\,j = \{1, 2, \ldots, n\}
\end{split}\label{eqn:cstr_chg_station_precedence}\\
&\sum_{j \in  \mathcal{L}^p} x_{v^k j}^k \le 1, \hspace{127pt}\forall k \in \mathcal{K} \label{eqn:cstr_agent_leave_hub}\\
&x_{v^k h}^k + \sum_{i \in \mathcal{L} \cup \mathcal{F} \setminus\{h\}} x_{ih}^k = \sum_{j \in \mathcal{L} \setminus\{h\}} x_{h j}^k, \hspace{1pt}\forall h\in \mathcal{L}^p, k \in \mathcal{K}\label{eqn:cstr_flow_continuity_lp}\\
&\sum_{i \in \mathcal{L} \setminus\{h\}} x_{i h}^k = \sum_{j \in \mathcal{L} \cup \mathcal{F} \cup \mathcal{H}_f \setminus\{h\}} x_{h j}^k, \hspace{14pt}\forall h\in \mathcal{L}^d, k \in \mathcal{K} \label{eqn:cstr_flow_continuity_ld}\\
&\sum_{i \in \mathcal{L}^d} x_{i h}^k = \sum_{j \in \mathcal{L}^p \cup \mathcal{H}_f} x_{h j}^k, \hspace{57pt}\forall h\in \mathcal{F}, k \in \mathcal{K} \label{eqn:cstr_flow_continuity_f}
\end{align}
\endgroup

%
%
\subsection{Timing Constraints}

Constraint \eqref{eqn:cstr_time_first_pickup} propagates the availability times from $\mathcal{H}_0$ to $\mathcal{L}^p$. Equation \eqref{eqn:cstr_time_continuity_l_to_l} guarantees time continuity across locations including request service times, and \eqref{eqn:cstr_time_p_before_d} ensures that pickup occurs before delivery. The mutually exclusive soft constraints \eqref{eqn:cstr_tw_pickup} and \eqref{eqn:cstr_tw_delivery} enforce the desired TW at either pickup or delivery, according to $\varrho^r$. We enforce temporal continuity from $\mathcal{L}^d$ to $\mathcal{F}$ with \eqref{eqn:cstr_time_continuity_ld_to_f} and from $\mathcal{F}$ to $\mathcal{L}^p$ with \eqref{eqn:cstr_time_continuity_f_to_lp}, which also incorporates the agent service and charging times. The succession condition \eqref{eqn:cstr_time_f_precedence} prevents simultaneous visits to a station and its duplicates, wherein a duplicate station is entered only after the agent service and charging times at its predecessor have elapsed. Equations \eqref{eqn:cstr_time_max_duration_ld} and \eqref{eqn:cstr_time_max_duration_f} compute the mission duration when arriving at $\mathcal{L}^d$ and $\mathcal{F}$, respectively, where excluding route segments to $\mathcal{H}_f$ evaluates the OVRP variant. Finally, we restrict each agent's trip time with \eqref{eqn:cstr_time_agent_workday} and determine the maximum mission duration with \eqref{eqn:cstr_time_mission_duration}.
\begingroup
\allowdisplaybreaks
\begin{align}
&t_j \ge \sum_{k \in \mathcal{K}} \left(\delta^k + c_{v^k j} \right) x_{v^k j}^k, \hspace{78pt}\forall j \in \mathcal{L}^p\label{eqn:cstr_time_first_pickup}\\
\begin{split}
&t_j \ge t_i + s^{\Gamma(i)} + c_{ij} - M \Bigl(1 - \sum_{k \in \mathcal{K}}x_{ij}^k\Bigr),\\
&\hspace{160pt}\forall i,j \in \mathcal{L},i \neq j
\end{split}\label{eqn:cstr_time_continuity_l_to_l}\\
&t_{d^r} \ge t_{p^r} + s^r, \hspace{128pt}\forall r \in \mathcal{R} \label{eqn:cstr_time_p_before_d}\\
&\underline{t}^r - \tau_{p^r} \le t_{p^r} \le \overline{t}^r + \tau_{p^r}, \hspace{82pt}\forall r \in \mathcal{R} \label{eqn:cstr_tw_pickup}\\
&\underline{t}^r - \tau_{d^r} \le t_{d^r} \le \overline{t}^r + \tau_{d^r}, \hspace{82pt}\forall r \in \mathcal{R} \label{eqn:cstr_tw_delivery}\\
\begin{split}
&t_j \ge t_i + s^{\Gamma(i)} + c_{ij} - M \Bigl(1 - \sum_{k \in \mathcal{K}} x_{ij}^k\Bigr),\\
&\hspace{161pt}\forall i \in \mathcal{L}^d, j \in \mathcal{F} 
\end{split} \label{eqn:cstr_time_continuity_ld_to_f}\\
\begin{split}
&t_j \ge t_i + \sum_{k \in \mathcal{K}} S^k x_{ij}^k + \sum_{l \in \{1,2,3\}} \xi_{i,l} + c_{ij}\\
&\hspace{16pt}- M \Bigl(1 - \sum_{k \in \mathcal{K}} x_{ij}^k\Bigr),\hspace{57pt}\forall i \in \mathcal{F}, j \in \mathcal{L}^p
\end{split} \label{eqn:cstr_time_continuity_f_to_lp}\\
\begin{split}
&t_{f_i^j} \ge t_{f_i^{j-1}} + \sum_{k \in \mathcal{K}} S^k \sum_{h \in \mathcal{L}^d} x_{h f_i^{j-1}}^k + \sum_{l \in \{1,2,3\}} \xi_{f_i^{j-1},l}\\
&\hspace{16pt}- M \Bigl( 2 - \sum_{k \in \mathcal{K}} \sum_{h \in \mathcal{L}^d} \bigl[x_{h f_i^j}^k + x_{h f_i^{j-1}}^k\bigr] \Bigr),\\
&\hspace{56pt}i = \{0, 1, \ldots, m-1\},~j = \{1, 2, \ldots, n\}
\end{split} \label{eqn:cstr_time_f_precedence}\\
\begin{split}
&T^k \ge t_i + s^{\Gamma(i)} + \sum_{j \in \mathcal{H}_f} c_{ij} x_{ij}^k - M \Bigl( 1 - \sum_{j \in \mathcal{H}_f} x_{ij}^k \Bigr),\\
&\hspace{160pt}\forall i \in \mathcal{L}^d, k \in \mathcal{K}
\end{split} \label{eqn:cstr_time_max_duration_ld}\\
\begin{split}
&T^k \ge t_i + S^k + \sum_{l \in \{1,2,3\}} \xi_{i,l} + \sum_{j \in \mathcal{H}_f} c_{ij} x_{ij}^k\\
&\hspace{16pt}- M \Bigl(1 - \sum_{j \in \mathcal{H}_f} x_{ij}^k \Bigr), \hspace{55pt}\forall i \in \mathcal{F}, k \in \mathcal{K}
\end{split} \label{eqn:cstr_time_max_duration_f}\\
&T^k \le D^k, \hspace{147pt}\forall k \in \mathcal{K} \label{eqn:cstr_time_agent_workday}\\
&T \ge T^k, \hspace{153pt}\forall k \in \mathcal{K} \label{eqn:cstr_time_mission_duration}
\end{align}
\endgroup

%
%
\subsection{Capacity Constraints} \label{ssb:cstr_capacity}
When traveling from $\mathcal{H}_0$ or $\mathcal{F}$ to $\mathcal{L}^p$, the agent's number of passengers is bounded by \eqref{eqn:cstr_load_psg_h0_f_to_lp}. Equation \eqref{eqn:cstr_load_psg_continuity_l_to_l} ensures load continuity between locations, \eqref{eqn:cstr_no_load_psg_to_f_hf} permits visits from $\mathcal{L}^d$ to $\mathcal{F}$ and $\mathcal{H}_f$ only with an empty load (no passengers), and \eqref{eqn:cstr_load_psg_limits} enforces the capacity limits. We impose the same set of restrictions on the equipment load with \eqref{eqn:cstr_load_eqp_h0_f_to_lp}--\eqref{eqn:cstr_load_eqp_limits}. Moreover, \eqref{eqn:cstr_load_combined_limits} models the configurable capacity, in which agents have convertible seats that can accommodate either seated passengers or stored equipment. $\gamma^k \ge 1$ indicates that one unit of equipment is equivalent to one or more passengers, and $\widetilde{Q}^k = Q_1^k + \gamma^k Q_2^k$ is used as the lowest upper bound to deactivate this constraint.
\begingroup
\allowdisplaybreaks
\begin{align}
\begin{split}
&u_{j,1}^k \ge q_1^{\Gamma(j)} - Q_1^k \Bigl(1 - x_{v^k j}^k - \sum_{i \in \mathcal{F}} x_{ij}^k \Bigr),\\ 
&\hspace{158pt}\forall j \in \mathcal{L}^p, k \in \mathcal{K} 
\end{split}\label{eqn:cstr_load_psg_h0_f_to_lp}\\
\begin{split}
&u_{j,1}^k \ge u_{i,1}^k + \mu_j q_1^{\Gamma(j)} - Q_1^k \Bigl(1 - x_{ij}^k \Bigr),\\
&\hspace{131pt}\forall i,j \in \mathcal{L}, i \neq j, k \in \mathcal{K}
\end{split}\label{eqn:cstr_load_psg_continuity_l_to_l}\\
&u_{i,1}^k \le Q_1^k \Bigl(1 - \sum_{j \in \mathcal{F} \cup \mathcal{H}_f} x_{ij}^k \Bigr), \hspace{38pt}\forall i \in \mathcal{L}^d, k \in \mathcal{K}\label{eqn:cstr_no_load_psg_to_f_hf}\\
&0 \le u_{i,1}^k \le Q_1^k, \hspace{100pt}\forall i \in \mathcal{L}, k \in \mathcal{K} \label{eqn:cstr_load_psg_limits}\\
\begin{split}
&u_{j,2}^k \ge q_2^{\Gamma(j)} - Q_2^k \Bigl(1 - x_{v^k j}^k - \sum_{i \in \mathcal{F}} x_{ij}^k \Bigr),\\ 
&\hspace{158pt}\forall j \in \mathcal{L}^p, k \in \mathcal{K} 
\end{split}\label{eqn:cstr_load_eqp_h0_f_to_lp}\\
\begin{split}
&u_{j,2}^k \ge u_{i,2}^k + \mu_j q_2^{\Gamma(j)} - Q_2^k \Bigl(1 - x_{ij}^k \Bigr),\\
&\hspace{131pt}\forall i,j \in \mathcal{L}, i \neq j, k \in \mathcal{K}
\end{split}\label{eqn:cstr_load_eqp_continuity_l_to_l}\\
&u_{i,2}^k \le Q_2^k \Bigl(1 - \sum_{j \in \mathcal{F} \cup \mathcal{H}_f} x_{ij}^k \Bigr), \hspace{38pt}\forall i \in \mathcal{L}^d, k \in \mathcal{K}\label{eqn:cstr_no_load_eqp_to_f_hf}\\
&0 \le u_{i,2}^k \le Q_2^k, \hspace{100pt}\forall i \in \mathcal{L}, k \in \mathcal{K} \label{eqn:cstr_load_eqp_limits}\\
\begin{split}
&u_{i,1}^k \le - \gamma^k u_{i,2}^k + Q_1^k + \widetilde{Q}^k \Bigl(1 - \sum_{j \in \mathcal{L} \setminus \{i\}} x_{ij}^k\Bigr),\\
&\hspace{160pt}\forall i \in \mathcal{L}^p, k \in \mathcal{K}
\end{split}\label{eqn:cstr_load_combined_limits}
\end{align}
\endgroup

\cite{c0082} suggested visiting $\mathcal{F}$ with passengers on board to decrease the overall mission duration. However, we argue that this type of detours detrimentally affects the customer experience and, therefore, excluded it.

%
%
\subsection{State-of-Charge (SoC) and Charging Constraints}
Equation \eqref{eqn:cstr_soc_first_pickup} determines the SoC at the first visited $\mathcal{L}^p$, and \eqref{eqn:cstr_soc_continuity_l_to_l} models the SoC drop between $\mathcal{L}$, including load-dependent discharge. Similarly, \eqref{eqn:cstr_soc_continuity_ld_to_f_hf} models the SoC change when traveling from $\mathcal{L}^d$ to $\mathcal{F}$ or $\mathcal{H}_f$, which occurs with zero occupancy as ensured by \eqref{eqn:cstr_no_load_psg_to_f_hf} and \eqref{eqn:cstr_no_load_eqp_to_f_hf}. Using \eqref{eqn:cstr_soc_min_leave_chg}--\eqref{eqn:cstr_soc_max_chg}, we compute the charging times based on the difference between the arrival and desired SoC, then the final SoC is further propagated with \eqref{eqn:cstr_soc_flow_chg}. Constraint \eqref{eqn:cstr_soc_limits} enforces the operational limits and prevents mid-route battery depletion. To ensure that agents always have sufficient energy after route completion, we recommend including route segments to $\mathcal{H}_f$ for both open and closed VRPs. We maintain the integrity of our battery model with \eqref{eqn:cstr_soc_min_enter_chg}--\eqref{eqn:cstr_soc_seg_done_order} as follows: first, \eqref{eqn:cstr_soc_min_enter_chg} ensures arriving at $\mathcal{F}$ with $\phi_i^k \le 0.85$, so that the subsequent constraints hold. Second, we link the binary completion variables $z_{i,1}, z_{i,2}$ with the charging times in the first and second segments using \eqref{eqn:cstr_soc_chg_fst_seg_ub}--\eqref{eqn:cstr_soc_chg_fst_seg_lb} and \eqref{eqn:cstr_soc_chg_snd_seg_ub}--\eqref{eqn:cstr_soc_chg_snd_seg_lb}, respectively. Third, \eqref{eqn:cstr_soc_max_fst_seg} and \eqref{eqn:cstr_soc_max_snd_seg} determine the maximum charging times upon completion of the previous segments for the second and third segments, respectively. Finally, \eqref{eqn:cstr_soc_seg_done_order} ensures the correct sequence of segment activation.
\begingroup
\allowdisplaybreaks
\begin{align}
&\phi_j^k \le \sigma^k - \alpha_0 c_{v^k j} + 1 - x_{v^k j}^k, \hspace{32pt}\forall j \in \mathcal{L}^p, k \in \mathcal{K}\label{eqn:cstr_soc_first_pickup}\\
\begin{split}
&\phi_j^k \le \phi_i^k - \alpha_0 c_{ij} - \alpha_1 c_{ij} u_{i,1}^k -  \alpha_2 c_{ij} u_{i,2}^k + 1 - x_{ij}^k,\\
&\hspace{131pt}\forall i, j \in \mathcal{L}, i \ne j, k \in \mathcal{K} 
\end{split}\label{eqn:cstr_soc_continuity_l_to_l}\\
\begin{split}
&\phi_j^k \le \phi_i^k - \alpha_0 c_{ij} + 1 - x_{ij}^k, \\
&\hspace{106pt}\forall i \in \mathcal{L}^d, j \in \mathcal{F} \cup \mathcal{H}_f, k \in \mathcal{K}
\end{split}\label{eqn:cstr_soc_continuity_ld_to_f_hf}\\
\begin{split}
&\phi_i^k \ge \widetilde{\sigma}^k - \sum_{l \in \{1,2,3\}} \beta_l \xi_{i,l} - 1 + \sum_{j \in \mathcal{L}^p \cup \mathcal{H}_f} x_{ij}^k,\\
&\hspace{162pt}\forall i \in \mathcal{F}, k \in \mathcal{K}
\end{split}\label{eqn:cstr_soc_min_leave_chg}\\
\begin{split}
&\phi_i^k \le 1 - \sum_{l \in \{1,2,3\}} \beta_l \xi_{i,l} + 1 - \sum_{j \in \mathcal{L}^p \cup \mathcal{H}_f} x_{ij}^k, \\
&\hspace{162pt}\forall i \in \mathcal{F}, k \in \mathcal{K}
\end{split}\label{eqn:cstr_soc_max_chg}\\
\begin{split}
&\phi_j^k \le \phi_i^k + \sum_{l \in \{1,2,3\}} \beta_l \xi_{i,l} - \alpha_0 c_{ij} + 1 - x_{ij}^k,\\
&\hspace{106pt}\forall i \in \mathcal{F}, j \in \mathcal{L}^p \cup \mathcal{H}_f, k \in \mathcal{K}
\end{split}\label{eqn:cstr_soc_flow_chg}\\
&\underline{\sigma}^k \le \phi_i^k \le 1, \hspace{63pt}\forall i \in \mathcal{L} \cup \mathcal{F} \cup \mathcal{H}_f, k \in \mathcal{K} \label{eqn:cstr_soc_limits}\\
&\phi_j^k \le 0.85 + 1 - \sum_{i \in \mathcal{L}^d} x_{ij}^k, \hspace{54pt}\forall j \in \mathcal{F}, k \in \mathcal{K} \label{eqn:cstr_soc_min_enter_chg}\\
&\phi_j^k \le 0.85 - \beta_1 \xi_{j,1} + 1 - \sum_{i \in \mathcal{L}^d} x_{ij}^k, \hspace{18pt}\forall j \in \mathcal{F}, k \in \mathcal{K} \label{eqn:cstr_soc_chg_fst_seg_ub}\\
&\phi_j^k \ge 0.85\hspace{1pt}z_{j,1} - \beta_1 \xi_{j,1} - 1 + \sum_{i \in \mathcal{L}^d} x_{ij}^k, \hspace{2pt}\forall j \in \mathcal{F}, k \in \mathcal{K} \label{eqn:cstr_soc_chg_fst_seg_lb}\\
\begin{split}
&\phi_j^k \le 0.95 - \beta_1 \xi_{j,1} - \beta_2 \xi_{j,2} + 1 - \sum_{i \in \mathcal{L}^d} x_{ij}^k,\\
&\hspace{162pt}\forall j \in \mathcal{F}, k \in \mathcal{K}
\end{split}\label{eqn:cstr_soc_chg_snd_seg_ub}\\
\begin{split}
&\phi_j^k \ge 0.95~z_{j,2} + 0.85 \left(z_{j,1} - z_{j,2}\right) - \beta_1 \xi_{j,1} - \beta_2 \xi_{j,2}\\
&\hspace{20pt}- 1 + \sum_{i \in \mathcal{L}^d} x_{ij}^k, \hspace{76pt}\forall j \in \mathcal{F}, k \in \mathcal{K}
\end{split}\label{eqn:cstr_soc_chg_snd_seg_lb}\\
&\beta_2 \xi_{j,2} \le 0.1 z_{j,1} + 1 - \sum_{k \in \mathcal{K}} \sum_{i \in \mathcal{L}^d} x_{ij}^k, \hspace{43pt}\forall j \in \mathcal{F} \label{eqn:cstr_soc_max_fst_seg}\\
&\beta_3 \xi_{j,3} \le 0.05 z_{j,2} + 1 - \sum_{k \in \mathcal{K}} \sum_{i \in \mathcal{L}^d} x_{ij}^k, \hspace{38pt}\forall j \in \mathcal{F} \label{eqn:cstr_soc_max_snd_seg}\\
&z_{i,2} \le z_{i,1}, \hspace{147pt}\forall i \in \mathcal{F} \label{eqn:cstr_soc_seg_done_order}
\end{align}
\endgroup

%
%
\subsection{Further Adaptations}

To highlight the flexibility of our model, we present some changes that readily extend it to different problems. We can model the multi-depot requirement of VRPs by forcing each agent to terminate its route at a certain hub $h^k \in \mathcal{H}_f$, with $\sum_{i \in \mathcal{L}^d \cup \mathcal{F}} x_{i h^k}^k = 1, \forall k \in \mathcal{K}$. To negate the need for tuning the objective values, we can set $y_r=1$ for certain requests to ensure their acceptance, regardless of priority. Finally, we may incorporate the prior occupancy of a station with $t_i \ge \omega_i, i \in \widetilde{\mathcal{F}}^0$, where $\omega_i$ is its time of earliest availability.

%
%
%

\section{Numerical Experiments}
\begin{figure}[htb]
	\centering
	\includegraphics[width=0.45\textwidth]{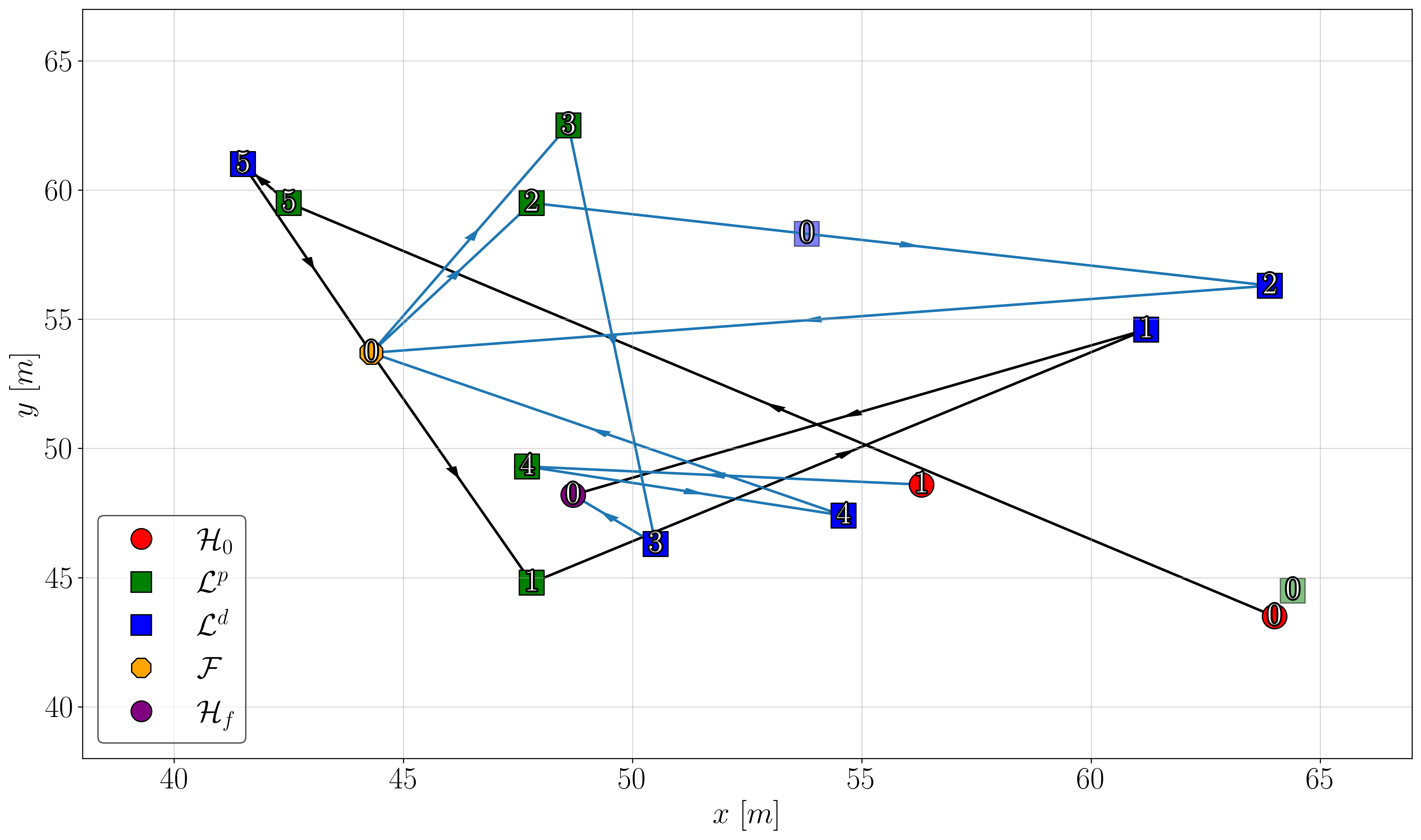}
	\caption{Selective (closed) VRP with $|\mathcal{K}| = 2$, $|\mathcal{R}| = 6$}
	\label{fig:test_scenario_1_routes}
\end{figure}

We implemented our approach with Pyomo (\cite{c0090}) and solved it with the academic license of Gurobi (\cite{c0095}). For brevity, we present two tests here and refer the reader to \textcolor{blue}{\url{https://doi.org/10.5281/zenodo.17704161}}, where the system specs, the code, and a total of 23 test scenarios with their computed solutions are available for reproducibility and transparency. The first scenario displayed in Fig. \ref{fig:test_scenario_1_routes} has $|\mathcal{K}| = 2$, $|\mathcal{R}| = 6$ and is solved as a selective VRP to optimality in $4.22[s]$. The computed routes of the first and second agents are depicted using black and blue arrows, respectively, and are:
\begin{itemize}
	\item \textbf{$k=0$}: $[v^0,~p^5,~d^5,~f_0^0,~p^1,~d^1,~h_0]$
	\item \textit{$k=1$}:  $[v^1,~p^4,~d^4,~f_0^1,~p^2,~d^2,~f_0^2,~p^3,~d^3,~h_0]$
\end{itemize}
where we used artificially high battery discharge rates to force multiple charging station visits. The first request $(r=0)$ was rejected and its nodes are thereby transparent; the selection and order of requests depend on their respective priorities $\lambda^r$ and the objective parameters $(\epsilon, \zeta, \eta)$. Table \ref{tbl:test_scenario_1_data} summarizes the relevant results, with minutes as the time unit. We also note the temporal consistency of charging station visits with $t_{f_0^0} = 41.4$, $t_{f_0^1}=52$, and $t_{f_0^2}=107.9$.

For the second test, we use typical battery parameters to solve a non-selective OVRP with $|\mathcal{K}| = 3$, $|\mathcal{R}| = 8$. Due to the problem size, we present the best achieved solution in $120[s]$ with a duality gap of $82\%$. The agents' routes are:
\begin{itemize}
	\item \textbf{$k=0$}: $[v^0,~p^4,~p^1,~d^1,~d^4,~p^7,~d^7]$
	\item \textit{$k=1$}:  $[v^1,~p^3,~p^2,~d^3,~d^2,~p^5,~d^5]$
	\item \textit{$k=2$}:  $[v^2,~p^0,~p^6,~d^0,~d^6]$
\end{itemize}
which highlights the Pickup-Pickup aspect of our approach.

\begin{table}[htb]
	\caption{First test scenario data}
	\label{tbl:test_scenario_1_data}
	\begin{supertabular*}{0.49\textwidth}{c || c | c | c | c | c || c | c | c | c | c }
		$r$ & $q_1^r$ & $q_2^r$ & $\varrho^r$ & $\underline{t}^r$ & $\overline{t}^r$ &$y_r$ & $k$ & $t_{p^r}$ & $t_{d^r}$ & $\tau$\\
		\midrule
		0 & 4 & 2 & $p$ & 38.9 & 49.1 & $\times$ & - & - & - & -\\
		1 & 1 & 2 & $d$ & 40.6 & 53.2 & $\checkmark$ & 0 & 61.6 & 80.4 & 27.2\\
		2 & 2 & 1 & $d$ & 43.4 & 55.3 & $\checkmark$ & 1 & 69.0 & 86.8 & 31.5\\
		3 & 3 & 2 & $d$ & 29.9 & 49.1 & $\checkmark$ & 1 & 130.2 & 149.1 & 100\\
		4 & 4 & 2 & $d$ & 14.4 & 34.4 & $\checkmark$ & 1 & 17.5 & 27.5 & 0.0\\
		5 & 2 & 2 & $p$ & 30.0 & 41.0 & $\checkmark$ & 0 & 27.0 & 31.2 & 3.0\\
	\end{supertabular*}
\end{table}
\begin{figure}[htb]
	\centering
	\includegraphics[width=0.45\textwidth]{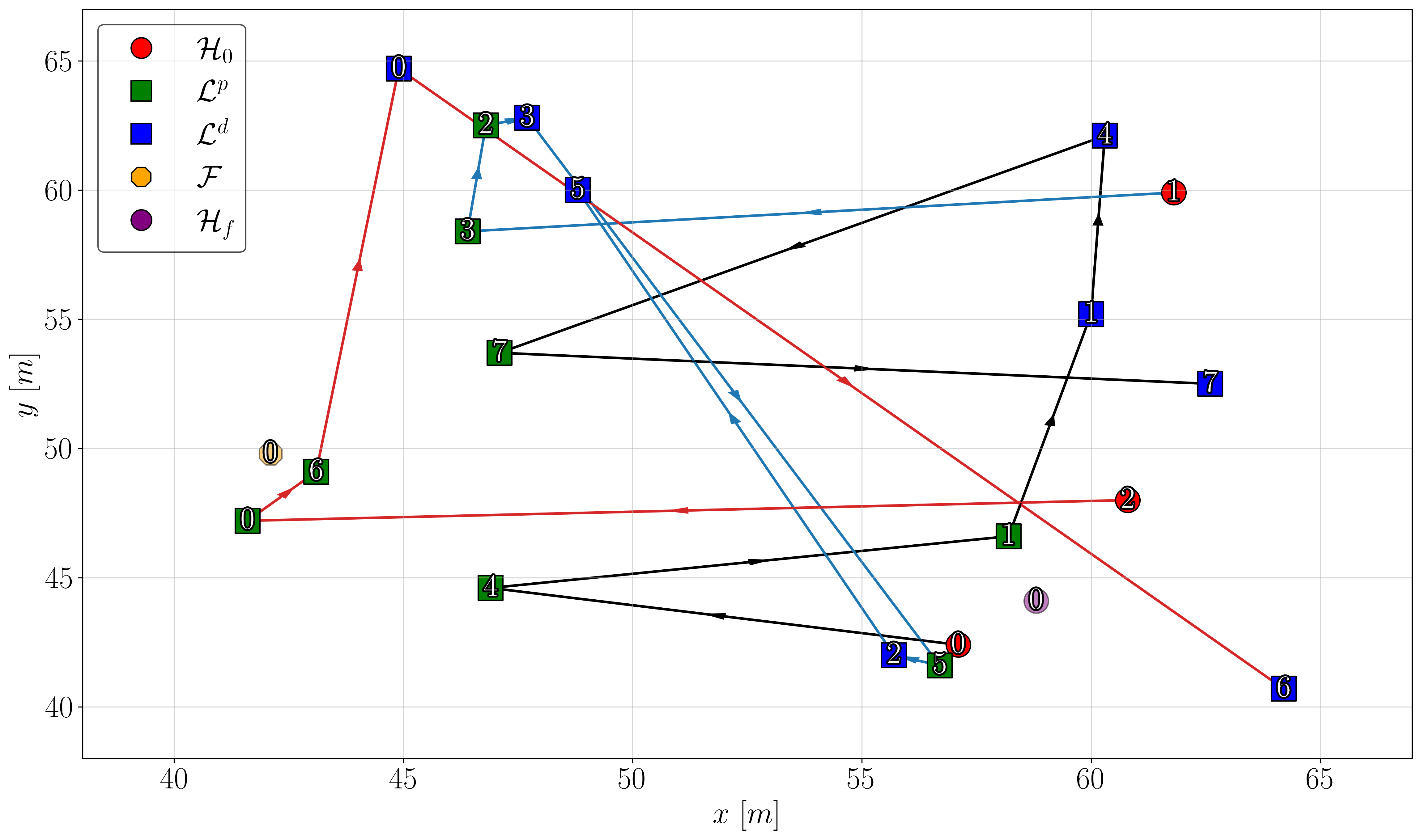}
	\caption{Non-selective OVRP with $|\mathcal{K}| = 3$, $|\mathcal{R}| = 8$}
	\label{fig:test_scenario_2_routes}
\end{figure}
\begin{table}[htb]
	\caption{Second test scenario data}
	\label{tbl:test_scenario_2_data}
	\begin{supertabular*}{0.49\textwidth}{c || c | c | c | c | c || c | c | c | c | c }
		$r$ & $q_1^r$ & $q_2^r$ & $\varrho^r$ & $\underline{t}^r$ & $\overline{t}^r$ &$y_r$ & $k$ & $t_{p^r}$ & $t_{d^r}$ & $\tau$\\
		\midrule
		0 & 2 & 1 & $p$ & 15.2 & 29.1 & $\checkmark$ & 2 & 24.7 & 54.7 & 0.0\\
		1 & 1 & 0 & $p$ & 20.9 & 32.4 & $\checkmark$ & 0 & 26.0 & 35.3 & 0.0\\
		2 & 3 & 1 & $d$ & 56.5 & 71.7 & $\checkmark$ & 1 & 25.7 & 67.0 & 0.0\\
		3 & 3 & 0 & $d$ & 41.5 & 57.8 & $\checkmark$ & 1 & 21.0 & 41.5 & 0.0\\
		4 & 2 & 1 & $d$ & 30.7 & 48.4 & $\checkmark$ & 0 & 22.0 & 42.7 & 0.0\\
		5 & 4 & 0 & $p$ & 49.6 & 66.8 & $\checkmark$ & 1 & 65.1 & 88.9 & 0.0\\
		6 & 4 & 1 & $p$ & 41.1 & 55.6 & $\checkmark$ & 2 & 37.2 & 87.9 & 3.96\\
		7 & 4 & 1 & $p$ & 56.1 & 73.3 & $\checkmark$ & 0 & 60.7 &79.1 & 0.0\\
	\end{supertabular*}
\end{table}

%
%
%

\section{Conclusion}
In this paper, we introduced the EM-DARP, a mobility-oriented VRP variant that augments the EV-DARP with agents of configurable capacities. We discussed the problem attributes, its implementation, and some extensions for real-life problems. We presented preliminary results and plan to develop more efficient solution methods for larger instances in the future, e.g., using heuristics and metaheuristics.

\bibliography{references}

\end{document}